%% file: template.tex
\pdfoutput=1
\documentclass{article}
\usepackage{arxiv}
\usepackage{amsmath,amsfonts,amsthm}
\usepackage[utf8]{inputenc} 
\usepackage[T1]{fontenc}    
\usepackage{hyperref}       
\usepackage{url}            
\usepackage{booktabs}       
\usepackage{amsfonts}       
\usepackage{nicefrac}       
\usepackage{microtype}      
\usepackage{lipsum}		
\usepackage{graphicx}
\usepackage{doi}
\usepackage{orcidlink}
\usepackage{siunitx}
\usepackage{subfig}
\usepackage{xcolor}
\usepackage{float}
\definecolor{myNewColorA}{RGB}{29,104,174}
\newtheorem{theorem}{Theorem} 
\newtheorem{definition}{Definition} 
\newtheorem{lemma}{Lemma} 
\newtheorem{corollary}{Corollary} 
\graphicspath{{fig/}}
\hypersetup{
colorlinks=true,
linkcolor=black,
citecolor=myNewColorA, 
}

\makeatletter
  \newcommand\figcaption{\def\@captype{figure}\caption}
  \newcommand\tabcaption{\def\@captype{table}\caption}
\makeatother

\title{Measuring Discrete Sensing Capability for ISAC via Task Mutual Information}

\author{ Fei~Shang$^{\orcidlink{0000-0002-5495-8869}}$ \\
	University of Science and Technology of China\\
	\texttt{shf\_1998@outlook.com} \\
\And
	Haohua Du$^*$$^{\orcidlink{0000-0002-8492-3990}}$ \\
	Beihang University\\
	\texttt{duhaohua@buaa.edu.cn} \\
 \And
 Panlong Yang$^*$$^{\orcidlink{0000-0003-1057-2793}}$ \\
	Nanjing University of Information Science and Technology\\
	\texttt{plyang@ustc.edu.cn} \\
 \And
 Xin He$^{\orcidlink{0000-0002-0125-4171}}$ \\
	Nanjing University of Information Science and Technology\\
	\texttt{xin.he@nuist.edu.cn} \\
 \And
	Jingjing Wang$^{\orcidlink{0000-0003-3170-8952}}$ \\
	Beihang University\\
	\texttt{drwangjj@buaa.edu.cn} \\
 \And
	Xiang-Yang Li$^{\orcidlink{0000-0002-6070-6625}}$ \\
	University of Science and Technology of China\\
	\texttt{xiangyangli@ustc.edu.cn} 
}

\renewcommand{\shorttitle}{Measuring Discrete Sensing Capability for ISAC \\via Task Mutual Information}

\hypersetup{
pdftitle={Towards the limits: Sensing Capability Measurement for ISAC Through Channel Encoder},
pdfsubject={CS.NI},
pdfauthor={Fei Shang, Haohua Du, Panlong Yang, Xin He, Wen Ma, Xiang-Yang Li},
pdfkeywords={ISAC, sensing channel, wireless sensing, mutual information},
}

\begin{document}
\maketitle

\begin{abstract}
	6G technology offers a broader range of possibilities for communication systems to perform ubiquitous sensing tasks, including health monitoring, object recognition, and autonomous driving.
Since even minor environmental changes can significantly degrade system performance, and conducting long-term posterior experimental evaluations in all scenarios is often infeasible, it is crucial to perform a priori performance assessments to design robust and reliable systems.
In this paper, we consider a discrete ubiquitous sensing system where the sensing target has \(m\) different states \(W\), which can be characterized by \(n\)-dimensional independent features \(X^n\).
The sensing algorithm extracts features from the received signals to obtain an embedding \(Y^n\), and then uses an appropriate decoding algorithm to infer the sensing result \(\hat{W}\) from \(Y_n\).
We define the \textit{discrete task mutual information} (DTMI) as \(I(X^n, Y^n)\), and consider the sensing to be in error when \(W \neq \hat{W}\).
We demonstrate that the expected lower bound of the sensing error $P_E^n$ for such systems is \([H(W)-I(X^n;Y^n)-H(P_E^n)]/\log m\), and the upper bound is \(\varepsilon + \sum_{k=1}^m p(w_k) \sum_{j\neq k}^m 2^{3n\varepsilon-\sum_{i=1}^n I(X_i(w_j);Y_i(w_k))}\), where \(H(\cdot)\) denotes the information entropy, and \(p(\cdot)\) represents the probability.
This model not only provides the possibility of optimizing the sensing systems at a finer granularity and balancing communication and sensing resources, but also provides theoretical explanations for classical intuitive feelings (like more modalities and more accuracy) in wireless sensing.
Furthermore, we validate the effectiveness of the proposed channel model through real-case studies, including person identification, displacement detection, direction estimation, and device recognition. The evaluation results indicate a Pearson correlation coefficient exceeding 0.9 between our task mutual information and conventional experimental metrics (e.g., accuracy). The open source address of the code is: \url{https://github.com/zaoanhh/DTMI}
\end{abstract}


\input{sections/intro-new}
\input{sections/model}

\input{sections/corollary}

\input{sections/case_study}

\input{sections/related_work}

\input{sections/conclusion}

\bibliographystyle{unsrt}
\bibliography{zotero}  






\end{document}

%% file: sections/intro-new.tex
\section{Introduction}
Thanks to its ubiquity, using radio frequency (RF) signals for sensing has found widespread applications.
In traditional integrated sensing and communication systems, such as joint radar-communication systems, common sensing tasks include target localization and tracking.
Recently, increasingly intelligent systems, such as smart agriculture, low-altitude economy, and smart healthcare, have demanded more comprehensive and continuous information sensing capabilities to support higher-level decision-making.
Traditional sensor-based sensing methods often provide only single-point measurements, making it difficult to achieve spatially continuous sensing.
Vision-based solutions, while powerful, struggle in low-light conditions and cannot provide temporally continuous sensing.
RF sensing, on the other hand, has the potential to offer both spatial and temporal continuity, meeting the multi-dimensional sensing needs of these intelligent systems.
Consequently, numerous advanced systems have been proposed, expanding the application scope of RF sensing to be more pervasive, including discrete state ubiquitous sensing task (such as material identification~\cite{wang2017tagscan,wuMSenseMobileMaterial2020}, and action recognition~\cite{adibCapturingHumanFigure2015,shangTameraContactlessCommodity2023}), and continuous state ubiquitous sensing task (such as health monitoring~\cite{yangArtificialIntelligenceenabledDetection2022,huangBreathLiveLivenessDetection2018}).
With the advent of the 6G era, it is anticipated that the sensing potential of RF systems will be further unleashed~\cite{saadVision6GWireless2020}.

However, many advanced sensing systems rely solely on field experiment results as the primary performance evaluation metric. 
Due to the significant impact that even minor environmental differences can have on sensing system performance, and the impracticality of conducting long-term experiments in real-world scenarios, while experimental evaluation is crucial, relying solely on post-hoc experimental methods hinders the development of robust and reliable system designs.
For example, the research by Chen et al.~\cite{chenTamingInconsistencyWiFi2017} demonstrated that simply opening a window, a minor environmental change, can cause the accuracy of indoor localization algorithms to drop by 80\%.
In many potential 6G signal application scenarios, such as vehicle-to-everything (V2X) communications~\cite{liuJointRadarCommunication2020} and autonomous driving~\cite{cuiVILAMInfrastructureassisted3D}, conducting long-term experiments to cover all possible environments is often impractical due to cost or ethical considerations.
Therefore, to design a robust and reliable sensing system, we first need to establish a priori performance metrics, much like complexity is to algorithms and channel capacity is to communications.

Traditionally, the system sensing capability usually be evaluated by analyzing how the received signals reflect the channel status, such as sensing mutual information $I(\mathbf{H};\mathbf{Y})$~\cite{zhang2021overview}, where $\mathbf{Y}$ is the received signal, and $\mathbf{H}$ is the channel status.
However, due to the following two reasons, it cannot fully adapt to the performance evaluation of ubiquitous sensing systems.
First, it is difficult to obtain complete information about the signal itself, we can only identify the sensory objects by analyzing several received signal features, such as the time-of-arrival (TOA), angle-of-arrival (AOA) and received signal strength (RSS). The relationship between the sensing capability of such features and the signal itself is ambiguous.
For example, when containing the same level of noise, the orientation difference of antennas may lead to an AoA (Angle of Arrival) estimation error exceeding tenfold~\cite{tai2019toward}.
Second, frequently the sensed objects have various types, including moving entities placed in the channel, temperature or humidity fields affecting the channel, etc.
The sensing capability analysis must be designed to operate for all possible types, not just the one that will actually be chosen since this is unknown at the time of design.

In this paper, we propose a general sensing channel encoder model to help determine the sensing capability of discrete ubiquitous sensing system -- the upper bound and lower bound of error in restoring the sensed object from given wireless signal features.
We consider a system performing \textbf{discrete sensing tasks}.
The target has $m$ different states, denoted as $W$.
We design $n$ independent features, $X^n$, to characterize these states.
After acquiring the received signal, we extract these $n$ features from the signal with sensing algorithm, which are represented as $Y^n$.
Finally, we use a decoding algorithm to obtain the sensing result, $\hat{W}$.
The discrete task mutual information (DTMI) $I(X^n;Y^n)$ is defined as the mutual information between $X^n$ and $Y^n$.
If the actual state of the target, $W$, differs from the sensing result, $\hat{W}$, we consider it an error.
In this paper, we demonstrate that the expected lower bound of the sensing error $p_E^n$ is 
\begin{equation}
    P_E^{n}\geq \frac{H(W)-I(X^n;Y^n)-H(P_E^n)}{\log m},
        \label{eq:lowbound}
\end{equation}
and the upper bound is
\begin{equation}
    P_E^{n}\leq \varepsilon + \sum_{k=1}^m p(w_k) \sum_{j\neq k}^m 2^{3n\varepsilon-\sum_{i=1}^n I(X_i(w_j);Y_i(w_k))},
    \label{eq:upbound}
\end{equation}
where $H(P_E^{n})=-P_E^{n}\log P_E^{n}-(1-P_E^{n})\log(1-P_E^{n})$, and $p(w_k)$ is the probability of the target being in state $w_k$.

\textbf{Main contributions:} 
(1) We propose a sensing channel encoder model to describe the sensing system, and derive the fundamental limits of specific sensing objects under given signal features, in terms of a performance measure called \textit{discrete task mutual information} (DTMI). This approach unifies such information from different features in a canonical form as a weighted sum associated with the weights characterizing the information intensity.

(2) Based on DTMI, we first provide upper and lower bounds of sensing errors for ubiquitous sensing systems and give a sufficient condition for lossless sensing. It enhances the interpretability of current sensing systems and can be further used to guide the problem of resource allocation for communication and sensing in ISAC systems.

(3) We validate the effectiveness of the proposed sensing system model in several real-world cases, including binary classification tasks such as Wi-Fi-based human identification and RFID-based displacement detection, and multi-classification tasks such as direction sensing based on electromagnetic signals and device identification based on traffic features. The experiment results show that the consistency between our proposed sensing capability evaluation method and the actual task results is up to 0.9 (Pearson correlation coefficient).

%% file: sections/model.tex
\section{Sensing channel encoder model}
\label{sec:model}
Sensing of discrete status finds broad applications in both industrial production and daily life scenarios, encompassing areas such as material identification, image recognition, and human presence detection.
In this section, we establish a discrete sensing channel encoder model to analyze the system's sensing capability.
Our analysis reveals that, with the status to be sensed being fixed, the DTMI directly dictates the lower and upper bounds of the expected sensing error.
Proceeding forward, we first introduce the definitions related to the discrete sensing channel encoder model, followed by an exploitation of DTMI to analyze the lower and upper bounds of the expected sensing error.
\begin{figure}[t]
	\centering
	\includegraphics[width=0.9\linewidth]{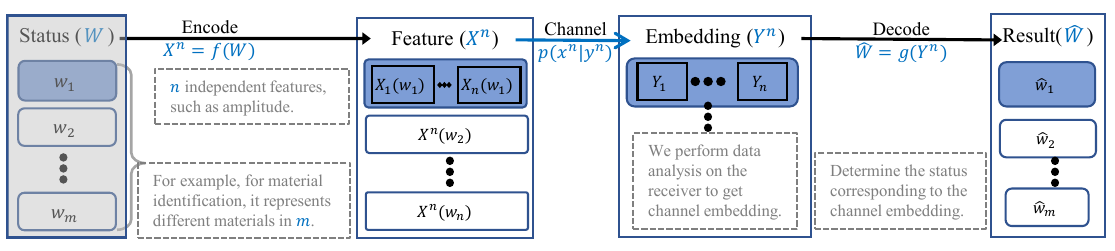}
	\caption{Sensing channel encoder.}
	\label{fig:model}
\end{figure}

\subsection{Model definitions.}
A typical sensing process often comprises several components: the target status to be sensed ($W$), the feature ($X^n$) designed to sense the status, the sensing channel embedding ($Y^n$) obtained through the sensing system, and the outcome ($\hat{W}$) derived after processing the signal.
We analyze the sensing system as shown in the Fig.~\ref{fig:model}.
The status $W$ has $m$ possible values, which together form the set $\mathcal{W}=\{w_1,\cdots,w_m\}$.
The probability that the target is in the $i$-th status is $\Pr(W=w_i)=p(w_i)$.
To facilitate the sensing of statuses, we construct $n$-dimensional independent features ${X}^n$ to represent the status $W$.
Given the status as $w_i$, the feature $X^n(w_i)$ is given by $X^n(w_i)=[X_1(w_i),\cdots, X_n(w_i)]$.
Upon transmission and subsequent data processing, the receiver is likely to receive this feature with a probability denoted as $p(y^n|x^n)$, which we represent as $Y^n$.
Subsequently, the receiver assesses the condition of the sensed target utilizing the acquired features $Y^n$ and decoding rules $g$.
The result is given by $\hat{W}=g(Y^n)$.
For instance, in a task of material identification using radio frequency (RF) signals, the targets possess varying materials ($W$).
We exploit the characteristic that different materials affect RF signals differently to design feature $X^n$, which are related to the amplitude of RF signals.
Then, using a receiver that captures electromagnetic waves in the space and processes them according to a sensing algorithm, we acquire the sensing channel embedding denoted as $Y^n$.
Finally, based on certain decision rules, we correlate $Y^n$ with the corresponding $X^n$ to ascertain the result $\hat{W}$.

To quantify the performance of the sensing system, we initially define the ``conditional error probability" and the ``expected value of the error".
The former represents the probability that the sensed result does not match the actual status $w_i$ given that the target status is $w_i$, while the latter signifies the expectation of the conditional error probabilities.
Furthermore, we introduce several definitions (Definition~\ref{def:independent_sequence},  \ref{def:joint}, and \ref{def:matching_set_joint}) to facilitate our analysis of the upper and lower bounds of the expected error value.

\begin{definition}
	\label{def:TMI}
	The \textbf{discrete task mutual information} (DTMI) is defined as the mutual information between the feature $X^n$ and the channel embedding $Y^n$, i.e., $I(X^n;Y^n)$.
\end{definition}

\begin{definition}
	\label{def:lambda_i}
	The \textbf{conditional error probability}  $\xi _i$ when the target status is $w_i$ is defined as:
	\begin{equation}
		\xi _i = \Pr(\hat{W}\neq w_i|W=w_i).
		\label{eq:lambda_i}
	\end{equation}
\end{definition}

\begin{definition}
	\label{def:error}
	The \textbf{expected value of the error}, denoted as $P_E^{n}$, is articulated as follows:
	\begin{equation}
		P_E^{n} = \sum_{i=1}^{m} p(w_i)\xi _i.
		\label{eq:error}
	\end{equation}
\end{definition}

\begin{definition}
	\label{def:independent_sequence}
	If a sequence \(X^n=[X_1,\cdots,X_n]\) of length \(n\), where each dimension is statistically independent of one another, we refer to sequence \(X^n\) as an \(n\)-dimensional \textbf{independent sequence}.
	Their joint probability density function is given by:
	\begin{equation}
		p(x^n)=\Pi_{i=1}^n p(x_i).
		\label{eq:independent_sequence}
	\end{equation}
\end{definition}
\begin{definition}
	\label{def:joint}
	For two \(n\)-dimensional independent sequences \(X^n\) and \(Y^n\), if the joint distribution of \((X^n,Y^n)\) is given by
	\begin{equation}
		p(x^n,y^n)=\Pi_{i=1}^np(x_i,y_i),
	\end{equation}
	we refer to \((X^n,Y^n)\) as a \(n\)-dimensional \textbf{jointly independent sequence}.
\end{definition}

\begin{definition}
	\label{def:matching_set_joint}
	The \textbf{jointly matching set} $B_{\varepsilon}^{(n)}$  of jointly independent sequence is defined as:
	\begin{equation}
		\begin{aligned}
			B_{\varepsilon}^{(n)} = & \left\{(X^n,Y^n)\in \mathcal{X}^n \times \mathcal(Y)^n: \right.                                            \\
			                        & \left|-\frac1n\mathrm{log}p(x^n)-\frac{1}{n}\sum_{i=1}^n H(X_i)\right|<\varepsilon                         \\
			                        & \left|-\frac1n\mathrm{log}p(y^n)-\frac{1}{n}\sum_{i=1}^n H(Y_i)\right|<\varepsilon                         \\
			                        & \left|\left.-\frac1n\mathrm{log}p(x^n,y^n)-\frac{1}{n}\sum_{i=1}^n H(X_i,Y_i)\right|<\varepsilon \right\}.
		\end{aligned}
		\label{eq:matching_set}
	\end{equation}
	where $(X^n,Y^n)$ is the \(n\)-dimensional jointly independent sequence.
 $H(X_i)$, $H(Y_i)$, and $H(X_i,Y_i)$ are the entropy of $X_i$, $Y_i$, and $(X_i,Y_i)$, respectively.
\end{definition}


\subsection{Lower bound on expected error.}
The current evaluation of sensing systems' performance predominantly relies on experimental assessments.
While experimental evaluations are highly effective in gauging system performance, conducting rigorous controlled experiments in real-world scenarios is exceedingly challenging.
Consequently, in many instances, it is difficult to ascertain whether the failure to achieve the desired accuracy is due to inadequately designed sensing features or simply unforeseen interference during the data acquisition process.
In this section, we give a lower bound on the expected error value based on DTMI, which helps us analyze the ultimate performance of the sensing system.

\begin{theorem}
	\label{thm:lowbound}
	For a sensing task $W$ with $m$ statuses, we use $n$ independent features to describe the status of the target.
	The expected value of the error $P_E^{n}$ satisfies the following lower bound:
	\begin{equation}
		P_E^{n}\geq \frac{H(W)-I(X^n;Y^n)-H(P_E^n)}{\log m}
		\nonumber
	\end{equation}
	where $H(P_E^n)=-P_E^n \log P_E^n - (1-P_E^n)\log(1-P_E^n)$.
\end{theorem}


\begin{proof}
	We first prove that the sensing model we defined forms a Markov chain.
	Then we combine Fano's inequality~\cite{tesunhanGeneralizingFanoInequality1994} and some properties of Markov chains to give a lower bound for $P_E^n$.
	\begin{lemma}
		For the sensing model described in Section~\ref{sec:model}, the target status $W$, the feature $X^n$, the received channel embedding $Y^n$, and the sensing result $\hat{W}$ form two Markov chains, i.e., $W\rightarrow X^n\rightarrow Y^n \rightarrow \hat{W}$ and $\hat{W}\rightarrow Y^n\rightarrow X^n\rightarrow W$.
  
		\begin{proof}
			For a Markov chain, some simple consequences are as follows~\cite{cover1999elements}:
			\begin{itemize}
				\item If $X\rightarrow Y\rightarrow Z$ is a Markov chain, $Z,Y,X$ form a Markov chain, i.e., $Z\rightarrow Y\rightarrow X$.
				\item For three random variables $X$, $Y$, and $Z$, if $Z=f(Y)$, then $X,Y,Z$ form a Markov chain, i.e., $X\rightarrow Y\rightarrow Z$.
			\end{itemize}
			According to the deification of sensing model, the feature is a function of the target status, i.e., $X^n=f(W)$;
			the sensing feature $Y^n$ is a function of the status feature $X^n$, i.e., $Y^n\sim p(y^n|x^n)$; and the sensing result $\hat{W}$ is a function of the sensing feature $Y^n$, i.e., $\hat{W}=g(Y^n)$.
			Therefore, the target status $W$, the feature $X^n$, the  channel embedding $Y^n$, and the sensing result $\hat{W}$ form a Markov chain, i.e., $W\rightarrow X^n\rightarrow Y^n \rightarrow \hat{W}$.
			Besides, we have $\hat{W}\rightarrow Y^n\rightarrow X^n\rightarrow W$.
		\end{proof}
	\end{lemma}
	According to the Fano's inequality~\cite{cover1999elements}, if three random variables $X,Y,Z$ form a Markov chain, i.e., $X\rightarrow Y \rightarrow Z$, we have:
	\begin{equation}
		\Pr\left(X\neq Z\right) \geq \frac{H(X|Z)-H(\Pr(X\neq Z))}{\log(|\mathcal{X}|)}.
		\label{eq:fano}
	\end{equation}
	where $H(X|Y)$ is the conditional entropy of $X$ given $Y$.
	For the Markov chain $W\rightarrow X^n\rightarrow Y^n \rightarrow \hat{W}$, according to the total probability formula and Ferno's inequality, we have:
	\begin{equation}
		\begin{aligned}
		P_E^{n} &= \Pr(\hat{W}\neq W) \geq \frac{H(W|\hat{W})-H(P_E^{n})}{\log(|\mathcal{W}|)} = \frac{H(W)-I(W;\hat{W})-H(P_E^{n})}{\log m}\\
		\end{aligned}
		\label{eq:lowbound1}
	\end{equation}

	According to the Data-processing inequality~\cite{cover1999elements}, if three random variables $X$, $Y$, and $Z$ form a Markov chain, $X\rightarrow Y\rightarrow Z$, then we have $I(X;Z)\leq I(X;Y)$, where $I(X;Y)$ is the mutual information between $X$ and $Y$.
	For the Markov chain $W\rightarrow X^n\rightarrow Y^n \rightarrow \hat{W}$, we have $I(W;\hat{W})\leq I(W;Y^n)$.
	And for the Markov chain $\hat{W}\rightarrow Y^n\rightarrow X^n\rightarrow W$, we have $I(Y^n;W)\leq I(Y^n;X^n)$.
	As a result, we have:
	\begin{equation}
		I(W;\hat{W}) \leq I(X^n;Y^n).
		\label{eq:lowbound1}
	\end{equation}

	Substituting Equ.~\eqref{eq:lowbound1} into Equ.~\eqref{eq:fano}, we have:
	\begin{equation}
		P_E^{n}\geq \frac{H(W)-I(X^n;Y^n)-H(P_E^n)}{\log m}.
  \nonumber
	\end{equation}
\end{proof}

\subsection{Upper bound on expected error.}
In communication, Shannon's second theorem~\cite{shannonMathematicalTheoryCommunication} posits that for a given signal, error-free transmission can always be achieved as long as we employ code words that are sufficiently long to encode the message.
This issue is equally pertinent in sensing: when the dimensionality $n$ of the feature is sufficiently large, what is the upper bound on the expected error? 
In this section, we derive an upper bound based on DTMI (Theorem~\ref{thm:upbound}) and provide a sufficient condition under which error-free sensing can be attained (Theorem~\ref{thm:main}).

\begin{theorem}
	\label{thm:upbound}
	For a sensing task with $m$ statuss, we use $n$ independent features to describe the status of the target.
	For sufficiently large $n$, the expected value of the error $P_E^{n}$ satisfies the following upper bound:
	\begin{equation}
		P_E^{n}\leq \varepsilon + \sum_{k=1}^m p(w_k) \sum_{j\neq k}^m 2^{3n\varepsilon-\sum_{i=1}^n I(X_i(w_j);Y_i(w_k))}
		\nonumber
	\end{equation}
\end{theorem}

\begin{proof}
	The expected error $P_E^n$ is influenced by the decision rule $g$, with the maximum likelihood criterion being a commonly employed rule in practical scenarios.
	However, for the sake of facilitating analysis, we introduce a novel decision rule defined in conjunction with the matching set $B_{\varepsilon}^{(n)}$ (Definition~\ref{def:matching_set_joint}), where in the result $\hat{W}$ is determined as $w_i$ whenever the channel embedding $Y^n$ and the feature $X^n(w_i)$ corresponding to the message $w_i$ form a jointly matching set.
	Under this rule, we first estimate the probability of $X^n, Y^n$ constituting a jointly matching set (Lemma~\ref{lem:matching1} to \ref{lem:matching3}) and subsequently present a suboptimal upper bound on the expected error (it is noted that employing alternative decision criteria might yield tighter upper bounds).
	
	\textbf{The decoding rule $g$}. To obtain sensing outcomes from $Y^n$, we employ the following rule \(g\):
	\begin{itemize}
		\item We declare that the target statue is $w_i$ if $(X^n(w_i),Y^n) \in B_{\varepsilon}^{(n)}$ and there is no other status $w_j$ such that $(X^n(w_j),Y^n) \in B_{\varepsilon}^{(n)}$.
		\item If there are multiple statuss $w_j$ such that $(X^n(w_j),Y^n) \in B_{\varepsilon}^{(n)}$ or there is no status $w_i$ such that $(X^n(w_i),Y^n) \in B_{\varepsilon}^{(n)}$, an error is declared.
	\end{itemize}

	To estimate the probability of an event occurring, we first prove the following lemma about matching sets.
	\begin{lemma}
		\label{lem:matching1}
		For a $n$-dimensional jointly independent sequence $(X^n,Y^n)$ and a matching set $B_{\varepsilon}^{(n)}$, when $n \to \infty$, the probability that $(X^n,Y^n)$ is in the matching set $B_{\varepsilon}^{(n)}$ is close to 1, which is
		\begin{equation}
			\Pr((X^n,Y^n)\in B_{\varepsilon}^{(n)}) \to 1.
			\label{eq:matching1}
		\end{equation}
	\end{lemma}
	\begin{proof}
		According to the Chebyshev's Law of Large Numbers, when the number of observations \(n\) is sufficiently large, the sample mean of \(n\) independent and identically distributed random variables converges in probability to their common expected value. Observing that the entropy is essentially the expectation of the logarithm of the reciprocal of probabilities, we leverage these two premises to underpin our proof.

		According to Chebyshev's Law of Large Numbers, given $\varepsilon>0$, there exists $n_1$, so that for all $n>n_1$, the following holds:
		\begin{equation}
			\begin{aligned}
				   P_1&=\Pr\left( \left|-\frac{1}{n}\log p\left(X^n\right)-\frac{1}{n}\sum_{i=1}^{n}H(X_i) \right| \geq \varepsilon\right)\\
				&=  \Pr\left( \left|\frac{1}{n}\sum_{i=1}^{n}\log p\left(X_i\right)-\frac{1}{n}\sum_{i=1}^{n}\mathbb{E}\left(\log p(X_i)\right) \right| \geq \varepsilon\right)<\frac{\varepsilon}{3}.
			\end{aligned}
		\end{equation}
		Similarly, there exists $n_2$ and $n_3$, so that for all $n>n_2$, the following holds:
		\begin{equation}
			\begin{aligned}
				 & P_2=\Pr\left( \left|-\frac{1}{n}\log p\left(Y^n\right)-\frac{1}{n}\sum_{i=1}^{n}H(Y_i) \right| \geq \varepsilon\right)<\frac{\varepsilon}{3}, \\
			\end{aligned}
		\end{equation}
		and for all $n>n_3$, the following holds:
		\begin{equation}
			\begin{aligned}
				 P_3=\Pr\left( \left|-\frac{1}{n}\log p\left(X^n,Y^n\right)-\frac{1}{n}\sum_{i=1}^{n}H(X_i,Y_i) \right| \geq \varepsilon\right)<\frac{\varepsilon}{3}.
			\end{aligned}
		\end{equation}
		Let $n_0=\max\{n_1,n_2,n_3\}$, then for all $n>n_0$, the following holds:
		\begin{equation}
			\begin{aligned}
				\Pr((X^n,Y^n)\in B_{\varepsilon}^{(n)}) > 1-(P_1+P_2+P_3) = 1-\varepsilon.
			\end{aligned}
		\end{equation}
	\end{proof}

	Going further, we consider the scenario where \((X^n, Y^n)\) forms a jointly independent sequence (Definition~\ref{def:joint}), and we examine the probability of them constituting a joint matching set.
	Initially, drawing upon Definition~\ref{def:matching_set_joint}, we estimate the counts of elements in both the matching set and the jointly matching set, which are related to the entropy.
	Specifically, the number of elements in the matching set for \(X^n\) and \(Y^n\) are approximately \(2^{\sum_{i=1}^n H(X_i)}\) and \(2^{\sum_{i=1}^n H(Y_i)}\), respectively, while the count of their joint matching sequences is roughly \(2^{\sum_{i=1}^n H(X_i,Y_i)}\).
	Building on this foundation, Lemma~\ref{lem:matching3} furnishes an estimate for the probability that \((X^n, Y^n)\) forms a joint matching set.

	\begin{figure}
	\begin{minipage}{0.47\linewidth}
		\centering
		\includegraphics[width=\linewidth]{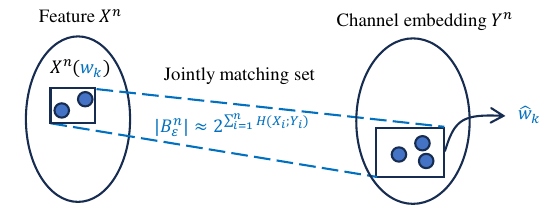}
		\caption{For independent sequences $X^n$ and $Y^n$, the number of elements in their jointly matching set is approximately $2^{\sum_{i=1}^n H(X_i,Y_i)}$. We decode the channel embedding as $\hat{w}_k$ when $Y^n$ forms a joint matching sequence with only one feature $X^n(w_k)$.}
		\label{fig:vn}
	\end{minipage}
	\hfill
	\begin{minipage}{0.47\linewidth}
		\centering
		\includegraphics[width=\linewidth]{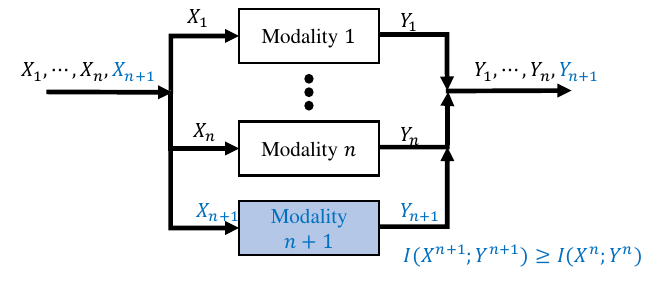}
		\caption{The sensing channel of multi-modal sensing system.}
		\label{fig:multimodal}
	\end{minipage}
\end{figure}
	\begin{lemma}
		\label{lem:matching2}
		The upper bound of the number of elements in the matching set of jointly independent sequence $B_\varepsilon^{(n)}$ is given by:
		\begin{equation}
			\Vert B_{\varepsilon}^{(n)}\Vert \leq 2^{n\varepsilon+\sum_{i=1}^n H(X_i,Y_i)},
			\label{eq:matching0}
		\end{equation}
		where $H(X_i,Y_i)$ is the entropy of $(X_i,Y_i)$, and $\Vert.\Vert$ denotes the number of elements in the set.
	\end{lemma}
	\begin{proof}
		According to the Definition~\ref{def:matching_set_joint}, if $(X^n,Y^n) \in B_{\varepsilon}^{(n)}$, we have:
		\begin{equation}
			p(x^n,y^n) \geq 2^{-n\varepsilon-\sum_{i=1}^n H(X_i,Y_i)}.
		\end{equation}
		As a result,
		\begin{equation}
			\begin{aligned}
				1 & = \sum_{(x^n,y^n) \in \mathcal{X}^n \times \mathcal(Y)^n} p(x^n,y^n) \geq \sum_{(x^n,y^n) \in B_{\varepsilon}^{(n)}} p(x^n,y^n)	 \geq 2^{-n\varepsilon-\sum_{i=1}^n H(X_i,Y_i)}|B_{\varepsilon}^{(n)}|.
			\end{aligned}
		\end{equation}
		Therefore, we have
		\begin{equation}
			\Vert B_{\varepsilon}^{(n)} \Vert  \leq 2^{n\varepsilon+\sum_{i=1}^n H(X_i,Y_i)}.\nonumber
   \nonumber
		\end{equation}
	\end{proof}
	\begin{lemma}
		\label{lem:matching3}
		For a $n$-dimensional jointly independent sequence $(\hat{X}^n,\hat{Y}^n)$ and a matching set $B_{\varepsilon}^{(n)}$, if $(\hat{X}^n,\hat{Y}^n)\sim p(x^n)p(y^n)$, i.e., $\hat{X}^n$ and $\hat{Y}^n$ are independent with the same marginals as $p(x^n,y^n)
		$, then
		\begin{equation}
			\Pr((\hat{X}^n,\hat{Y}^n)\in B_{\varepsilon}^{(n)}) \leq 2^{3n \varepsilon-\sum_{i=1}^n I(X_i;Y_i)},
			\label{eq:matching2}
		\end{equation}
		where $I(X_i;Y_i)$ is the mutual information between $X_i$ and $Y_i$.
	\end{lemma}

	\begin{proof}
 According to the definition of the jointly matching set, we have:
 \begin{equation}
     \begin{aligned}
         \log (p(x^n)) \leq n\varepsilon -\sum_{i=1}^n H(X_i),\ \text{and}\ 
         \log (p(y^n)) \leq n\varepsilon -\sum_{i=1}^n H(Y_i)
     \end{aligned}
 \end{equation}
		The probability of a joint independent sequence $(\hat{X}^n,\hat{Y}^n)$ in $B_\varepsilon^{n}$ is given by:
		\begin{equation}
			\begin{aligned}
					  \Pr((\hat{X}^n,\hat{Y}^n) \in B_{\varepsilon}^{(n)}) &= \sum_{(x^n,y^n) \in B_{\varepsilon}^{(n)}} p(x^n)p(y^n) \leq   \left|B_\varepsilon^{(n)}\right| 2^{n\varepsilon -\sum_{i=1}^n H(X_i)} 2^{n\varepsilon -\sum_{i=1}^n H(Y_i)}\\
				& \leq  2^{3n\varepsilon+\sum_{i=1}^n (H(X_i,Y_i)-H(X_i)-H(Y_i))} = 2^{3n \varepsilon-\sum_{i=1}^n I(X_i;Y_i)}.
			\end{aligned}
		\end{equation}
	\end{proof}
	
	We first estimate the probability that the sensing result $\hat{W}$ is wrong when the target status is $W=w_i$.
	We can assume without loss of generality that the target status is $w_1$.
	We consider the following events:
	\begin{equation}
		C_i=\left\{(X^n(w_i),Y^n(w_1)) \in B_{\varepsilon}^{(n)}\right\}, \quad i\in \{1,\cdots,m\}.
	\end{equation}
	where $Y^n(w_1)$ is the received channel embedding when the target status is $w_1$.
	Based on the decision rule and Definition~\ref{def:lambda_i}, the conditional error probability at this point is given by:
	\begin{equation}
		\begin{aligned}
			\xi _1 & = Pr\left(\bar{C_1}\bigcup_{i=2}^{m}C_i\right) \leq Pr\left(\bar{C_1}\right)+\sum_{i=2}^{m}Pr\left(C_i\right),
		\end{aligned}
		\label{eq:xi_1}
	\end{equation}
	where $\bar{C_1}$ is the complement of $C_1$.

	According to Lemma~\ref{lem:matching1}, we have:
	\begin{equation}
		\begin{aligned}
			\Pr\left(\bar{C_1}\right) \leq \varepsilon.
		\end{aligned}
	\end{equation}
	Besides, for $j\in \{2,\cdots, m\}$, the feature $X^n(w_j)$ is independent of $X^n(w_1)$, so is $X^n(j)$ and $Y^n(w_1)$.
	Hence, according to Lemma~\ref{lem:matching3}, we have:
	\begin{equation}
		\begin{aligned}
			\Pr\left(C_j\right) \leq 2^{3n\varepsilon-\sum_{i=1}^n I(X_i(w_j);Y_i(w_1))}.
		\end{aligned}
	\end{equation}
	Substituting the above results into Eq.~\eqref{eq:xi_1}, we have:
	\begin{equation}
		\begin{aligned}
			\xi _1 & \leq \varepsilon + \sum_{j=2}^m 2^{3n\varepsilon-\sum_{i=1}^n I(X_i(w_j);Y_i(w_1))}.
		\end{aligned}
		\label{eq:xi_2}
	\end{equation}
	According to Definition~\ref{def:error}, we have:
	\begin{equation}
		\begin{aligned}
			P_E^{n} & = \sum_{k=1}^{m} p(w_k)\xi _k\leq \varepsilon + \sum_{k=1}^m p(w_k) \sum_{j\neq k}^m 2^{3n\varepsilon-\sum_{i=1}^n I(X_i(w_j);Y_i(w_k))}.
		\end{aligned}
	\end{equation}
\end{proof}

Finally, Theorem~\ref{thm:main} provides a sufficient condition for error-free sensing, indicating that for achieving error-free sensing, a sufficient number of features with high DTMI must be identified~\footnote{This requirement diverges from the conclusion in communications, where merely having a sufficient number of codewords is typically sufficient.}.

\begin{theorem}
	\label{thm:main}
	For a sensing task with $m=2^{nR}$ statuss, we use $n$ independent features to describe the status of the target.
	For a sufficiently large $n$, if $R$ satisfies the following equation, 
 	\begin{equation}
		R < \min_{k \neq j} I (\bar{X}^n(w_k); \bar{Y}^n(w_j)) - 3\varepsilon,
		\label{eq:main}
	\end{equation}
	where $\bar{X}(w_j)$ and $\bar{Y}(w_j)$ is the mean $X^n(w_j)$ and $Y^n(w_j)$, we have $\xi_j  \rightarrow 0$.
\end{theorem}

\begin{proof}
	In Theorem~\ref{thm:upbound}, we derive an upper bound estimate for the expected error $P_E^n$.
	Capitalizing on the convexity property of mutual information, we leverage Jensen's inequality to provide a sufficient condition for a tight error estimation.
	This approach ensures that our estimate effectively captures the inherent relationship between the variables, harnessing the convexity to yield a more robust and accurate analysis of the error's expected magnitude without loss of generality.

According to the Jensen's inequality, if $f$ is a convex function and $X$ is a random variable, we have:
\begin{equation}
	f(\mathbb{E}(X)) \leq \mathbb{E}(f(X)).
\end{equation}
Since the mutual information is a convex function~\cite{cover1999elements}, we have:
\begin{equation}
	n I(\bar{X}^n;\bar{Y}^n) \leq n \sum_{i=1}^n \frac{1}{n}I(X_i;Y_i),
\end{equation}
where $\bar{X}^n$ and $\bar{Y}^n$ is the mean of $X^n$ and $Y^n$.
As a result, for a $j\in \{1,\cdots,m\}$, the Equ.~\eqref{eq:xi_1} can be rewritten as:
\begin{equation}
\xi_j \leq \varepsilon + \sum_{k\neq j}^m 2^{3n\varepsilon-n I(\bar{X}^n(w_k);\bar{Y}^n(w_j))}.
\end{equation}

As a result, for $m=2^{nR}$ and sufficiently large $n$, if $R$ satisfies the Equ.~\eqref{eq:main}, we have:
\begin{equation}
	\xi_j \leq \varepsilon + 2^{3n\varepsilon+nR-n \min_{k\neq j}I (\bar{X}^n(w_k); \bar{Y}^n(w_j))} \rightarrow 2\varepsilon.
\end{equation}
\end{proof}

%% file: sections/corollary.tex
\section{Corollary}
\label{sec:corollary}

Previous excellent sensing systems have summarized many valuable experiences, such as multi-modal systems tend to achieve better sensing performance. However, these experiences currently lack theoretical explainability. In this section, we employ sensing channel encoder model and DTMI as tools to attempt to explain some classic phenomena.


\subsection{Why do multimodal systems tend to exhibit superior performance?}
\label{sec:multimodal}
In a communication system, Shannon's second theorem stipulates that the error rate can be reduced to an arbitrary low level, provided that the codewords are sufficiently lengthy.
Similarly, many previous research works have shown that using multi-modality for sensing helps achieve better performance, which can be explained by the theorem we proved previously.
In this subsection, we will theoretically explain why multi-modal sensing systems are more capable of achieving superior sensing performance based on the DTMI.

\begin{figure}[t]
    \centering
    \includegraphics[width=0.9\linewidth]{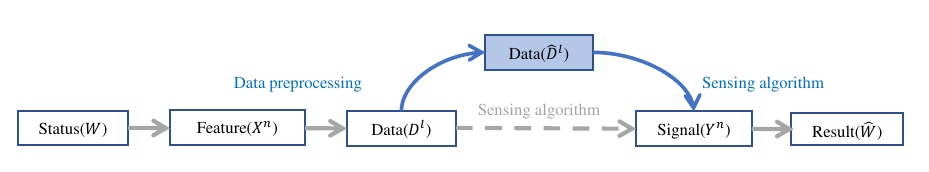}
    \caption{The sensing system with data pre-processing.}
    \label{fig:preprocessing}
\end{figure}
Fig.~\ref{fig:multimodal} shows a schematic diagram of a multi-modal system.
For the target state $W$, we use $n$ modalities to sense it.
The channels of different modalities are directly independent of each other.
For example, in order to identify the material of the target, we use three modalities: vision, sound wave, and radio frequency signal for sensing.
The transmission of visual signal, sound wave signal, and radio frequency signal is independent of each other.
According to the Theorem~\ref{thm:lowbound}, when the number of states $m$ remains unchanged, the lower bound of the expected value of the error $P_E^n$ is related to $I(X^n;Y^n)$.
Note that both mutual information and conditional mutual information are non-negative.
When we add a new mode, we have
\begin{equation}
   \begin{aligned}
    I(X^{n+1};Y^{n+1})  &= I(X^n,X_{n+1};Y^n,Y_{n+1}) \\
         &= I(X^n;Y^n) + I(X^n;Y_{n+1}|Y^n) + I(X_{n+1};Y^{n+1}|X^n) \\
             &\geq I(X^n;Y^n),
    \end{aligned}
\end{equation}
where $X^{n+1}=[X_1,X_2,\ldots,X_n,X_{n+1}]$ and $Y^{n+1}=[Y_1,Y_2,\ldots,Y_n,Y_{n+1}]$.
Therefore, the more modalities we use, the larger the mutual information $I(X^n;Y^n)$, the lower the theoretical lower bound of the expected value of the error.

\begin{figure*}[t]
	\centering
	\subfloat[]{\includegraphics[width=0.18\linewidth]{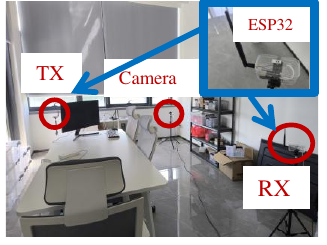}\label{fig:human-a}}
	\hfill
	\subfloat[]{\includegraphics[width=0.5\linewidth]{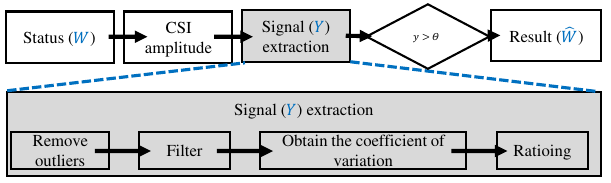}\label{fig:human-b}}
	\hfill
	\subfloat[]{\includegraphics[width=0.27\linewidth]{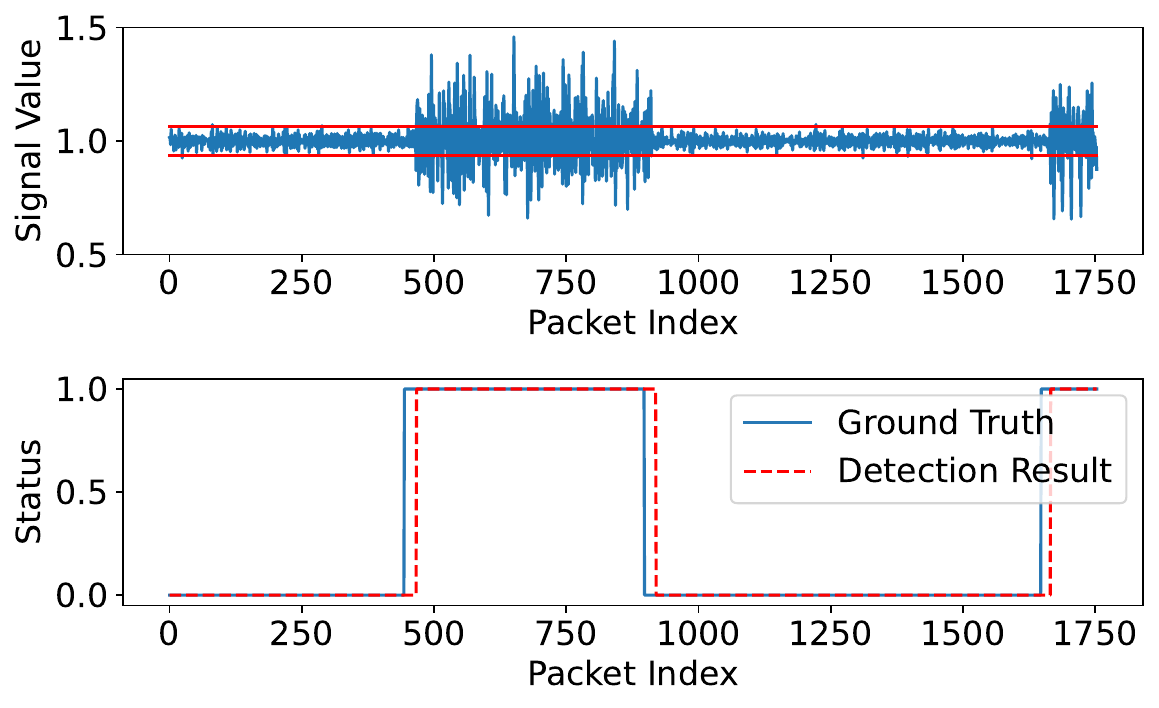}\label{fig:human-c}}
	\caption{Human detection in home environments based on WiFi. (a) Experimental environment and device deployment. (b) Human detection algorithm based on thresholding method. (c) Channel embedding extraction.}
	\label{fig:human}
\end{figure*}

\subsection{How do we compare which of two sensing features is better?}
In the process of designing a sensing system, it is crucial to carefully craft the sensing features.
To show that feature $X$ is better than feature $X'$, we usually need to run many micro-benchmarks.
While experimental validation is a compelling method of verification, it frequently involves intricate setup procedures and can be time-consuming.
Moreover, due to the challenge of deploying tests across a wide range of scenarios, it is often difficult to ascertain whether feature $X$ is truly superior to feature $X'$ or if this conclusion holds only in specific contexts.

In this paper, we propose DTMI which can reflect the performance of sensing features to a certain extent.
Specifically, we consider two features $X$ and $X'$.
After passing through the sensing channel, their corresponding channel embeddings are $Y$ and $Y'$, respectively.
According to Theorem~\ref{thm:lowbound} and Theorem~\ref{thm:upbound}, both the upper and lower bounds of the expected error are related to the DTMI.
If the DTMI $I(X;Y)>I(X';Y')$, the upper and lower bounds of the expected value of the error $P_E$ will be reduced, which means that it is easier to achieve good performance using $X$ as sensing features.
This necessitates alternative approaches, beyond experimental validation, to assess the performance of designed sensing features.

\subsection{Is data pre-processing a ``cure-all" solution?}

Since data contains a lot of noise and interference, sensing systems usually include a data preprocessing module when they are designed, which is used to improve data quality for subsequent processing.
Previous studies have shown that preprocessing can often improve sensing performance.
Now our questions are: can we accomplish any sensing task with arbitrary accuracy through sufficiently sophisticatedly designed data preprocessing algorithms?

We refine the sensing channel encoder model depicted in Fig.~\ref{fig:model}, and the result is illustrated in Fig.~\ref{fig:preprocessing}.
Specifically, for the $n$-dimensional independent features \(X^n\), after transmission through an actual physical channel, we obtain an $l$-dimensional data \(D^l\) at the receiver.
For instance, to localize a target using radio frequency (RF) signals, we employ angle of arrival (AoA) as a feature.
At the receiver, what we receive is the amplitude and phase of the RF signals, which are $D^l$.
Subsequently, we subject the received data \(D^l\) to data preprocessing, yielding a processed data \(\hat{D}^l\).
Then we utilize the sensing algorithm to process the data $\hat{D}^l$ to obtain the channel embedding $Y^n$, and finally use the judgment algorithm to obtain the result $\hat{W}$.
In particular, when no data preprocessing is used, it is equivalent to $\hat{D}^l=D^l$.

\begin{corollary}
    If the following equation holds,
    \begin{equation}
        H(W)-I(X^n;D^l) > 1,
        \label{eq:preprocessing}
    \end{equation}
    lossless sensing cannot be achieved simply by improving the effect of data preprocessing.
\end{corollary}
\begin{proof}
According to the definition Markov chain, the channel shown in Fig.~\ref{fig:preprocessing} constitutes a Markov chain $W\rightarrow X^n \rightarrow D^l \rightarrow \hat{D}^l \rightarrow Y^n \rightarrow \hat{W}$.
Note that ``whether the sensing result is correct" is a binary event, so we have $H(P_E^n)\leq 1$.
According to the Theorem~\ref{thm:lowbound} and the Data-Processing Inequality, we have
\begin{equation}
    \begin{aligned}
       P_E^n &\geq \frac{H(W)-I(X^n;Y^n)-H(P_E^n)}{\log m} \geq \frac{H(W)-I(X^n;D^l)-1}{\log m} \geq 0.
    \end{aligned}
\end{equation}
Therefore, lossless sensing cannot be achieved simply by improving the effect of data preprocessing.
\end{proof}

%% file: sections/case_study.tex
\begin{figure}
	\centering
	\includegraphics[width=0.5\linewidth]{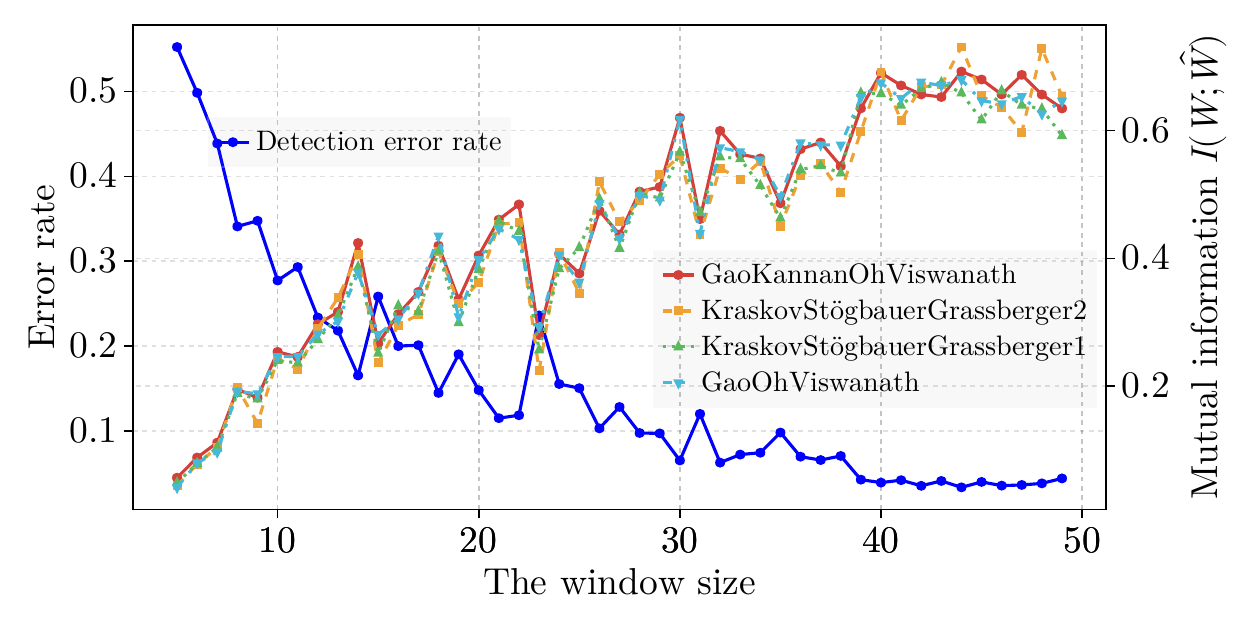}
	\caption{The accuracy exhibits a similar trend to the mutual information estimated by numerical algorithms.}
	\label{fig:person_MI}
\end{figure}

\section{Case Study}
\label{sec:case}

We illustrate the role of system performance evaluation based on sensing channel encoder model and DTMI through several case studies.
We begin by examining the application of DTMI in binary classification tasks, using examples of human detection in home settings via WiFi and appliance cabinet door displacement detection in industrial scenarios via RFID.
For multi-class classification, we consider two instances: the classic sensing problem in ISAC systems – direction estimation, and device identification based on an open-source traffic dataset.
The results demonstrate that across different cases, the Pearson correlation between the trend of DTMI changes and that of accuracy fluctuations exceeds 0.9.
Furthermore, DTMI can provide estimates of upper and lower bounds for sensing system errors, which is beneficial for optimizing and balancing ISAC systems.

\subsection{Binary classification task.}

(1) \textbf{Human detection based on WiFi devices.}

Indoor human detection plays a pivotal role in services such as elderly monitoring.
In particular, device-free passive human detection has garnered significant attention in recent years.
While methods based on infrared, pressure sensors, and the like have been applied to human detection, they either rely on specialized hardware or come at a higher cost.
Moreover, vision-based and infrared-based methods are only effective within line-of-sight (LOS) coverage.
Wi-Fi devices, being one of the most widely deployed radio frequency devices, have led to the implementation of numerous radio frequency sensing systems around them.
In recent years, with the advancement of wireless sensing technology, Wi-Fi-based approaches have proven to be a promising method for indoor human detection.
We deployed an experiment based on Wi-Fi devices in a residential setting and estimated mutual information using numerical methods.
The experimental results indicate that DTMI exhibits a similar trend to accuracy.
In this case study, their Pearson correlation coefficient exceeds $0.9$.

\begin{figure*}[t]
    \centering
    \subfloat[]{\includegraphics[width=0.15\linewidth]{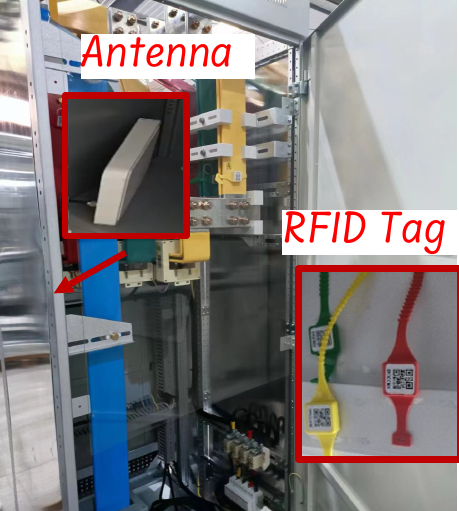} \label{fig:door_a}}
    \hfill
    \subfloat[]{\includegraphics[width=0.57\linewidth]{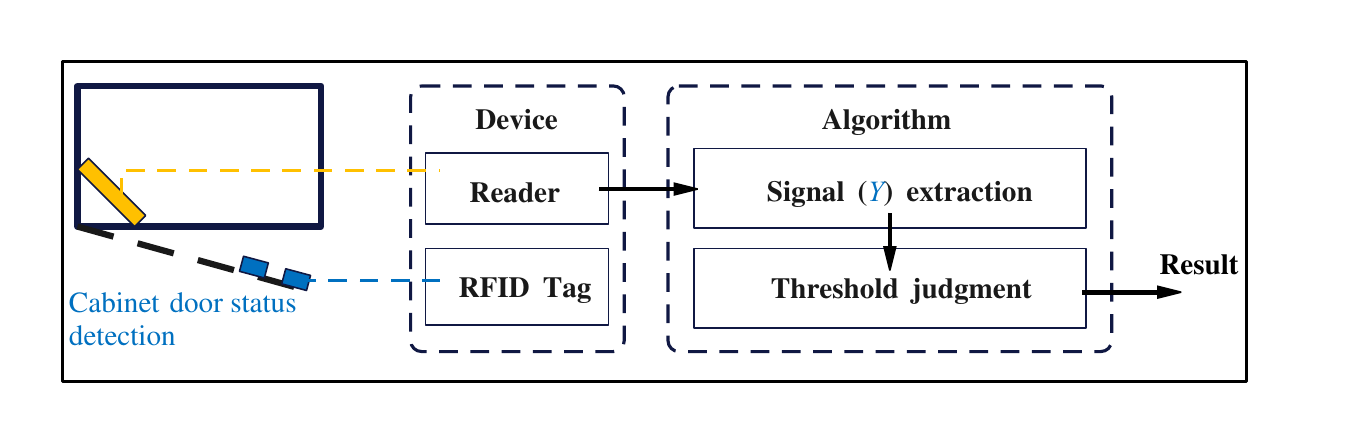} \label{fig:door_b}}
    \hfill
    \subfloat[]{\includegraphics[width=0.25\linewidth]{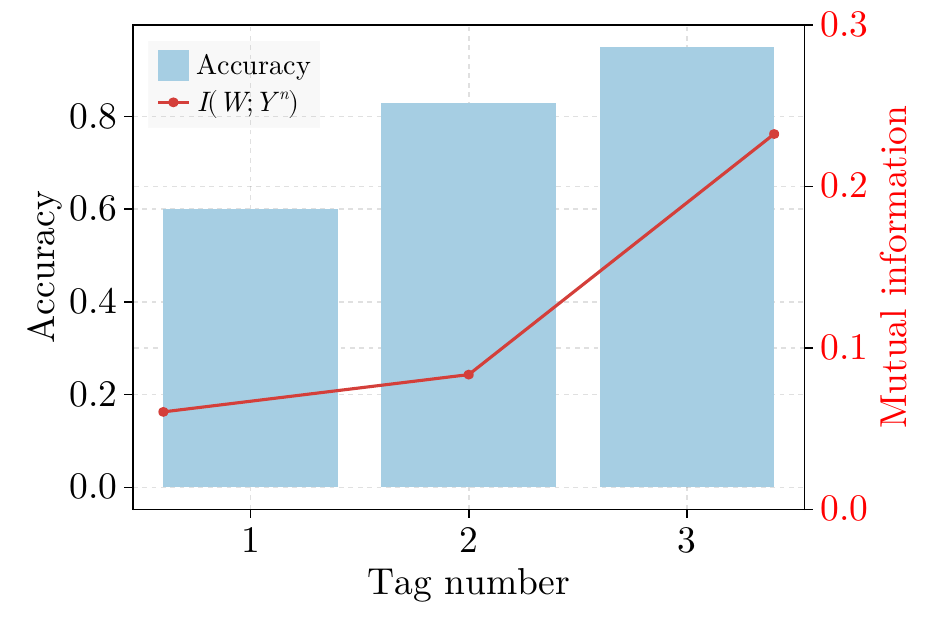} \label{fig:door_c}}
	\caption{RFID-based electrical cabinet door state monitoring. (a) Schematic diagram of device deployment. (b) The cabinet door status monitoring algorithm. (c) The identification accuracy (represented by the bar chart on the left y-axis) and the mutual information (indicated by the red line on the right y-axis) exhibit a consistent trend of variation.}
\end{figure*}
The experimental setup is depicted in Fig.~\ref{fig:human-a}, where we conducted experiments in a $\qty{4}{m} \times \qty{6}{m}$ office using an ESP32 device as both transmitter and receiver, each equipped with a single antenna.
Additionally, a camera was placed within the environment to capture video footage for recording ground truth.
The sampling rate of the ESP32 is set to \qty{100}{Hz}.
Ten volunteers are invited to participate in the tests.
Each data acquisition session lasted 10 minutes: the first 5 minutes ensured the room is empty, followed by 5 minutes with human activity (walking) inside the room.

State $W$ has two possible values: ``personnel present" and ``personnel absent".
After obtaining CSI data, we initially sliced the data, then performed data preprocessing to eliminate outliers and apply filtering.
Finally, channel embedding $Y$ is extracted from this processed data and compared against empirical thresholds to ascertain the presence or absence of individuals, which is the result $\hat{W}$.
The entire data processing procedure is illustrated in Fig.~\ref{fig:human-b}.
The corfficient of variation of $k$-th subcarrier is $\delta_{\Delta T}^k=\sigma_{\Delta T}^k/\mu_{\Delta T}^k$, where $\Delta T$ is the width of the time window, $\mu_{\Delta T}^k$ and $\sigma_{\Delta T}^k$ are the mean and standard deviation of the $k$-th subcarrier, respectively.
And the channel embedding $y$ is given by $y = \frac1n\sum_{i=1}^1\left|\delta^i\Delta_T/\delta^i\Delta_{T-1}\right|$, where $n$ is the number of subcarriers.
If $y$ falls within the experiential threshold range, we consider the environment to be ``person absent"; otherwise, it is determined to be person present.
The entire data processing workflow is illustrated in Fig.~\ref{fig:human-b}.
Here, the threshold range is $[0.935, 1.065]$.
Figure~\ref{fig:human-c} shows a example of the channel embedding extraction process.

In Fig.~\ref{fig:person_MI}, the blue solid line illustrates the error rate of human detection as the width of the time window varies $\Delta T$.
The dashed lines of other colors represent the mutual information $I(W;\hat{W})$ under different numerical estimation algorithms, namely KraskovStogbauerGrassberger1~\cite{kraskovEstimatingMutualInformation2004}, KraskovStogbauerGrassberger2~\cite{kraskovEstimatingMutualInformation2004}, GaoKannanOhViswanath~\cite{gaoEstimatingMutualInformation}, and GaoOhViswanath~\cite{gaoDemystifyingFixed$k$2018}.
The results demonstrate that the trend of accuracy change is highly consistent with the trend of mutual information change, indicating that in such tasks, DTMI can serve as an additional performance metric, complementing accuracy, to evaluate system performance.

(2) \textbf{RFID-based electrical cabinet door direction monitoring.}
Ensuring electrical safety is crucial during the manufacturing process.
Take the electrical cabinet as an example; if its door is inadvertently opened without timely detection, there are potential safety hazards, including the risk of electrical fires and electric shock.
In the field of terminal sensing in power systems, electromagnetic transformer-type sensors have traditionally dominated.
In recent years, non-electric quantity sensing technologies such as vibration, stroke, arc light, and spectral sensing have gained widespread application in digital electrical equipment and power systems.
However, these sensing technologies frequently depend on specialized sensors that boast high sensitivity and accuracy.
These sensors are typically burdened with several drawbacks, including complexities in power supply, large size and weight, high energy consumption, vulnerability to electromagnetic interference, difficult installation processes, and exorbitant costs. Consequently, they fall short of meeting the requirements for the development of modern smart power equipment.
Given the cost-effectiveness and ease of deployment of RFID tags, we have developed an algorithm for monitoring cabinet door status using multiple tags.
Furthermore, we employ the mutual information of tasks, as proposed in this paper, to assess the system’s performance.

We conduct relevant tests in a factory setting.
For an industrial metal electrical cabinet (measuring approximately $\qty{1}{m} \times \qty{1}{m}\times \qty{2}{m}$) used in production, our objective is to monitor the status of the cabinet door.
The RFID reader is ImpinJ Speedway R420 reader.
The RFID system operates in the $\qty{920}{MHz} \sim \qty{926}{MHz}$.
Two states $W$ are defined: when the door opening angle is less than $5^{\circ}$, it is considered ``closed"; otherwise, it is deemed ``open".
We affix several (1 to 3) anti-metal RFID tags onto the cabinet door and positioned the antenna within the cabinet body.
The deployment configuration of the equipment is illustrated in Fig.~\ref{fig:door_a}.
After collecting the RSSI (Received Signal Strength Indicator) from each tag, we perform differential processing against an initial value, followed by calculating the average of these differential values across multiple tags.
If the average differential exceeds an empirically determined threshold (set here as $2.5$), we conclude that the sensing result is ``open"; otherwise, it is concluded as ``closed".
The detailed steps of data processing are depicted in Fig.~\ref{fig:door_b}.

\begin{figure}[t]
    \centering
	\includegraphics[width=0.65\linewidth]{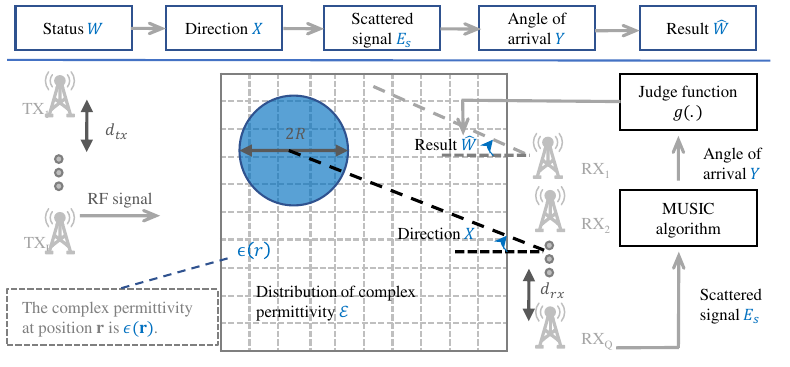}
	\caption{The sensing channel of direction estimation based on Music algorithm and electromagnetic signal.}
	\label{fig:aoa}
\end{figure}
The results of the state monitoring are shown in Fig.~\ref{fig:door_c}.
Due to the cabinet being made of metal, the electromagnetic waves suffer from severe multipath interference.
Consequently, when only one tag is used, the stability of the data is poor, and the empirical threshold becomes almost unusable after the tag position shifts by just a few centimeters.
This issue leads to an identification accuracy of less than 60\%.
This is well reflected by the mutual information $I(W;Y^n)$ ($n=1$), which has a small value in this case.
Since the spacing of the tags exceeds half a wavelength, their mutual influence is minimal, and thus we can approximately consider the reflection signals from different tags as independent of each other.
Consequently, following corollary introduced in Sec.~\ref{sec:multimodal}, as the number of tags increases, so does the mutual information.
We employ GaoOhViswanath~\cite{gaoDemystifyingFixed$k$2018} method to estimate the mutual information, and the red line in Fig.~\ref{fig:door_c} illustrates its trend, which increases with the number of tags.
As the mutual information increases, so does the accuracy of state identification.

\subsection{Multiple classification tasks.}
\begin{figure}
\begin{minipage}{0.47\linewidth}
\centering
	\includegraphics[width=\linewidth]{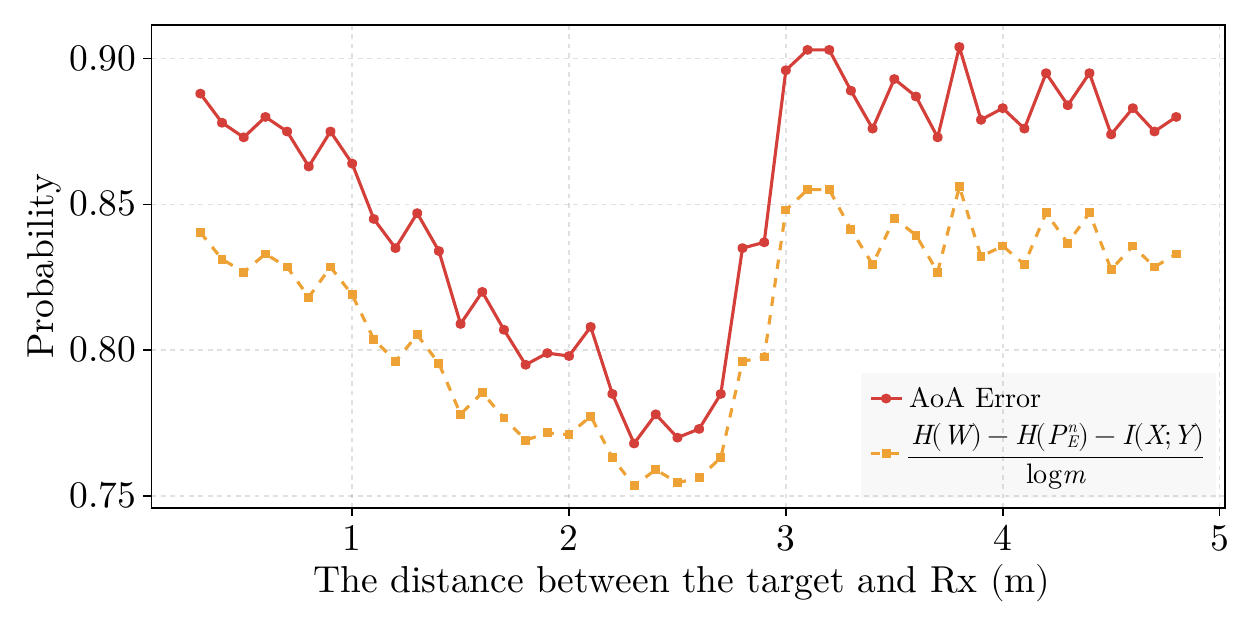}
	\caption{The sensing channel of AoA estimation based on Music algorithm and electromagnetic signal.}
	\label{fig:aoa_distance}
\end{minipage}
\hfill
\begin{minipage}{0.475\linewidth}
    \centering
       \includegraphics[width=\linewidth]{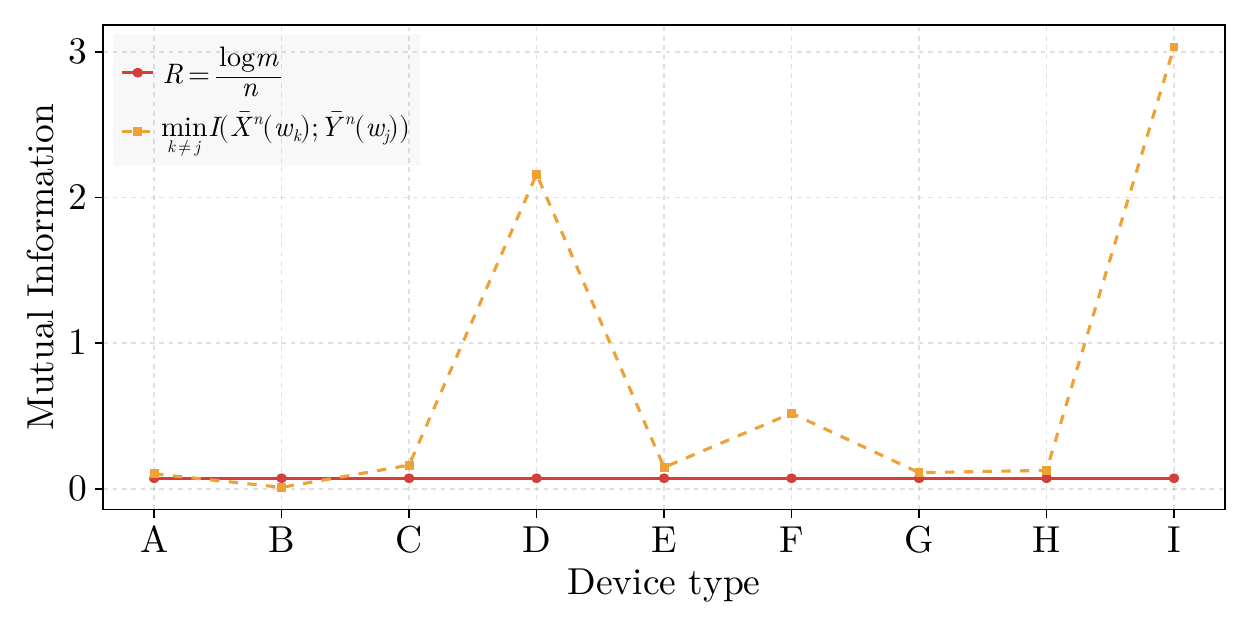}
       \caption{Device type identification based on traffic characteristics.}
       \label{fig:device}
   \end{minipage}
\end{figure}
(1) \textbf{Direction estimation based on Music algorithm and electromagnetic signal.}

Location sensing represents one of the most prevalent and fundamental tasks in the field.
A plethora of superior systems have been developed utilizing location sensing.
Nevertheless, for an extended period, there has been a dearth of methods other than experimental evaluations to assess the influence of numerous factors, including the distance between the target and both the transmitter and receiver, on localization accuracy.
In this case study, we use direction estimation based on the Music algorithm (one of the most popular localization algorithms)~\cite{kotaruSpotFiDecimeterLevel2015} and electromagnetic signal to show the application of the proposed framework.

We consider a two-dimensional direction estimation problem.
The basic model setup is shown in the Fig.~\ref{fig:aoa}.
There are $P$ transmitting antennas and the position of the $p$-th transmitting antenna is denoted as $\mathbf{r}_{tx_p}$.
The receiver has $Q$ receiving antennas and the position of the $q$-th receiving antenna is denoted as $\mathbf{r}_{rx_q}$.
The distance between two adjacent antennas is $d_{rx}$ and $d_{tx}$ for the receiver and transmitter, respectively.
The distribution of complex permittivity in space is $\mathcal{E}$, and the permittivity at position $\mathbf{r}$ is $\mathcal{E}=\epsilon(\mathbf{r})$.
For ease of calculation, we set the shape of the target to be a circle with a radius of $2R$.
We set $m$ states, each state corresponds to a direction interval.
The direction is defined as the angle (the $X$ in Fig.~\ref{fig:aoa}) between the line connecting the center of the target circle and the center of the receiving antenna array and the vertical line of the antenna array.
The direction interval is $[-\pi,\pi]$, which is evenly divided into $m$ sub-intervals.
The scattered signals $E_s$ are calculated using Maxwell's equations and the method of moments~\cite{shangLiquImagerFinegrainedLiquid2024}.
After adding Gaussian random noise to $E_s$, we estimate signal $ Y$ using the MUSIC algorithm.
Finally, we use the maximum likelihood algorithm to determine the direction $X$ corresponding to channel embedding $Y$, and then output the category to which $X$ belongs as the result $\hat{W}$.

We first simulated the effect of the distance between the target and the receiver on the direction estimation accuracy.
During the simulation, we set the parameters as follows.
We set the number of states $m=9$.
The frequency of the electromagnetic signal is \qty{5.0}{GHz}.
The distance between the transmitter and the receiver is \qty{8.0}{m}.
There are $P=1$ transmitting antennas and $Q=3$ receiving antennas.
The distance between two adjacent receiving antennas is \qty{0.03}{m}, i.e., $d_{rx}=\qty{0.03}{m}$.
The diameter of the target is $2R=\qty{0.2}{m}$.
The distance between the target and the receiver changes from \qty{0.3}{m} to \qty{5}{m}.
The material of the target is water, and the permittivity is given by empirical formula~\cite{kaatzeComplexPermittivityWater1989}.
In order to solve the scattered waves $E_s$ using the moment method, we discretized the space so that each subunit is a square with a side length of \qty{0.01}{m}.

We estimate the mutual information using a numerical algorithm~\cite{kristian_agasoster_haaga_2023_8409495}.
The results are shown in Fig.~\ref{fig:aoa_distance}.
The results show that when the target is too close to the receiver, the accuracy of the direction estimation is very poor.
We believe this is because the existence of phenomena such as diffraction makes it difficult to use the ray tracing model (the basic assumption of the MUSIC algorithm) to equivalent signal transmission~\cite{shang2022liqray}.
When the distance is too large, the accuracy will also decrease.
We believe this is because the scattered wave signal becomes weaker, resulting in a decrease in angular resolution.
In addition, the changing trend of accuracy is basically consistent with the changing trend of the error lower bound given by our DTMI, and their Pearson correlation coefficient exceeds 0.95.

(2) \textbf{Device type identification based on traffic characteristics.}

\begin{figure}[t]
    \centering
	\includegraphics[width=0.9\linewidth]{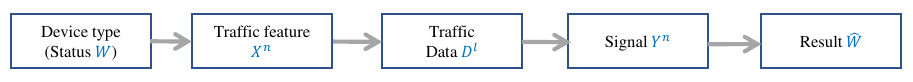}
	\caption{Device type identification based on traffic characteristics.}
	\label{fig:device_step}

\end{figure}
\begin{table}[t]
    \centering
    \footnotesize
    \caption{The code names for device types and their actual names.}
    \label{tab:device}

    \begin{tabular}{|l|l|}

    \hline
        \textbf{Device name} & \textbf{Device type} \\ \hline
        XIAOMI Bedside Lamp & A \\ \hline
        HUAWEI TC5206 & B \\ \hline
        XIAOMI Induction Cooker & C1, C2 \\ \hline
        HUAWEI Matebook & D \\ \hline
        XIAOMI Microwave Oven & E1, E2 \\ \hline
        Oneplus6T & F \\ \hline
        XIAOMI Rice Cooker & G1, G2 \\ \hline
        XIAOMI EPS & H1, H2, H3 \\ \hline
        XIAOMI Table Lamp & I1, I2 \\ \hline
    \end{tabular}
\end{table}
Security and privacy issues have always been a hot topic among researchers~\cite{han2022accuth}.
In recent years, with the development of the Internet of Things (IoT) and WiFi technology, attackers have devised more diverse means to steal private information.
For instance, many attackers place concealed cameras and other IoT devices designed to pilfer private information in public environments such as hotels. After acquiring this private information, these devices continuously transmit the data through gateways.
To detect illegal devices, Yan et al.~\cite{yan2022real} leveraged the characteristic that different devices generate distinct traffic patterns, using the traffic at the gateway for device type identification.
Their research findings indicated a minimum accuracy rate of 99.17\% for identifying common devices like various models of Xiaomi phones, routers, etc.
In this paper, based on their open-source code and data, our analysis shows that lossless detection can be achieved when the bit rate satisfies the sufficient condition given in Theorem~\ref{thm:main}.

At this moment, the schematic diagram illustrating the sensing channel encoder model is depicted in Fig.~\ref{fig:device_step}.
Post-processing of the traffic data, we employ the methodology put forth by Yan.~\cite{yan2022real} and colleagues to derive a 30-dimensional signal intended for appliance classification.
Our dataset encompasses traffic information from eleven distinct device categories, whose precise nomenclature and coding are presented in Table~\ref{tab:device}.
Notably, instances where identical device names are associated with multiple codes signify the existence of several units of the same device category.
As an illustration, Type ``C" comprises two devices, labeled ``C1" and ``C2", which denote two separate models of Xiaomi induction stoves.
The evaluation procedure incorporates a five-fold cross-validation strategy, alongside adopting the KNN classifier as the analytical tool for discrimination.
Throughout every iteration of cross-validation, the signals hailing from the subset earmarked for training are denoted as $X^n$, whereas those belonging to the testing subset are marked as $Y^n$, precedented by applying algorithm ``GaoOhViswanath"~\cite{gaoDemystifyingFixed$k$2018} to gauge mutual information.
Fig.~\ref{fig:device} illustrates the results of our calculations. Here, the possible state number $m =9$, and 30-dimensional features are used for device type recognition.
In this case, the corresponding sensing bitrate is $R = \log{m}/{n}$.
We find that the data at this time satisfies the sufficient conditions given by Theorem~\ref{thm:main}, and the goal of non-destructive sensing can be achieved at this time.
the DTMI values are mostly above the lower bound given by Theorem~\ref{thm:main}, and the overall recognition accuracy of KNN has exceeded 99\%.

%% file: sections/related_work.tex
\section{Related work}
\label{sec:related}

\subsection{Sensing systems based on communication devices.}
\label{subsec:sensing_systems}
ISAC is widely acknowledged as a pivotal enabler for a myriad of emerging applications, encompassing smart manufacturing, smart homes, and smart cities~\cite{saadVision6GWireless2020}.
The deployment of professional sensing equipment on a large scale is often impeded by their substantial size and high costs.In the context of the burgeoning Internet of Things (IoT), a multitude of endpoints, originally intended for communication purposes such as WiFi, speakers and microphones, RFID, among others, have gained prominence due to their abundance and cost-effectiveness in comparison to specialized equipment.
Consequently, a growing number of researchers and practitioners are exploring the use of these devices for sensing tasks. These applications range from localization and trajectory tracking to material identification and health monitoring.

(1) \textbf{Localization and trajectory tracking.}
The proliferation of wireless devices, coupled with the development of wireless network infrastructure, has led to a significant increase in their deployment within both workplaces and homes.
Recently, there has been a notable trend towards employing these communication devices for mobile trajectory tracking.
Wi-Fi based systems \cite{wang2018witrace,kotaru2017localizing} use Channel State Information/Received Signal Strength Indicator (CSI/RSSI) for localization~\cite{kotaru2017localizing,qian2017widar,ali2015keystroke}, gesture tracking \cite{wang2018witrace}, gesture recognition \cite{ali2015keystroke,tan2016wifinger,abdelnasser2015wigest}, etc.
Within this context, Widar \cite{qian2017widar} quantifies the relationship between CSI dynamic changes and user location and speed to achieve an average position error of \qty{25}{cm}.
RFID-based systems can achieve centimeter-level tracking accuracy using phase-based methods \cite{yang2014tagoram,adib2013see,wang2018multi,wang2016d}.
For instance, RF-IDraw\cite{wangRFIDrawVirtualTouch2015} utilizes interference techniques to measure the relative phase between multiple RFID readers, while Tadar~\cite{adib2013see} achieves through-wall tracking by exploiting multipath signal variations caused by human movement.
Tagoram \cite{yang2014tagoram} uses the concept of "virtual antennas" and phase holography to map measured phases to possible tag locations and calculates moving trajectories through phase changes.
However, the positioning accuracy of these works is influenced by many factors, such as the number of antennas and noise levels. There is still a lack of theoretical means to quantify the impact of these factors on the results.

(2) \textbf{Material identification.}
In contrast to professional equipment, which can cost tens of thousands of dollars, radio frequency signal transceivers are comparatively less expensive and compact.
This makes them more feasible for deployment in lightweight sensing scenarios such as homes, or large-scale scenarios like warehouses.
For instance, we can utilize commercial WiFi signals to detect whether the purchased fruits are ripe~\cite{tan2021object}.
In addition, compared with visible light, the frequency of wireless signals is lower, which makes them have better propagation performance in low light or non-line-of-sight environments.
For example, for liquids placed in opaque containers, many radio frequency signal-based systems can identify the solution concentration with a granularity of 1\% ~\cite{shang2022liqray,shangContactlessFineGrainedLiquid2024}.

(3) \textbf{Health monitoring.}
Both heartbeat and respiratory behavior produce corresponding body-conducted sounds.
Consequently, sound has emerged as a significant modality for the sensing of vital signs.
Xiong et al.~\cite{li2020fm} have extended the effective distance of acoustic sensing by utilizing ubiquitous sound waves, achieving accurate personnel tracking, gesture tracking, eye movement tracking, etc. in multiple scenarios.
For a long time, auscultation has been an important part of sleep and respiratory related research, so many works use microphones on mobile devices to capture the air-conducted sounds of respiration for snoring detection, and sleep apnea detection.
Han et al.~\cite{han2023breathsign} employ in-ear microphones to facilitate sense and user identity authentication via respiratory behavior analysis.
Owing to its non-contact characteristic, radio frequency signals have the potential to alleviate pressure on users during monitoring or sensing processes. 
This has led to a surge in academic interest in this field in recent years.
Wang et al.~\cite{wang2016human} first propose the Fresnel zone theory of WiFi signal sensing in free space, theoretically exploring the impact of human breathing depth on the reception of radio frequency signals.

However, unlike the performance of communication systems that can be reasonably assessed using theoretical metrics like channel capacity, the current evaluation of sensing system performance largely relies on experimental approaches.

\subsection{Performance measurement of the ISAC system.}
Traditional research often treats ``communication" and ``sensing" as two distinct systems.
However, a growing body of recent studies has demonstrated that these two concepts are intrinsically interconnected in the context of information theory, forming an intriguing ``odd couple"~\cite{blissCooperativeRadarCommunications2014}.
In recent years, a significant number of researchers have dedicated their efforts to examining the theoretical performance of systems through the lens of synesthesia.
A typical system modeling method is the linear Gaussian model, which is~\cite{xiong2023fundamental}
\begin{equation}
    \begin{aligned}
        \mathbf{Y} = \mathbf{H} \mathbf{X} + \mathbf{N},
    \end{aligned}
\end{equation}
where $\mathbf{Y}$ is the received signal, $\mathbf{X}$ is the transmitted signal, $\mathbf{H}$ is the channel matrix, and $\mathbf{N}$ is the noise matrix.
From the perspective of communication, the fundamental problem is how to accurately estimate the transmitted signal $\mathbf{X}$ from the received signal $\mathbf{Y}$.
In accordance with Shannon's second law, the ultimate performance of a channel is dictated by its capacity.
This capacity is intrinsically linked to the mutual information between $\mathbf{X}$ and $\mathbf{Y}$, denoted as $I(\mathbf{X};\mathbf{Y})$.
From the perspective of sensing, the basic problem is to estimate $\mathbf{H}$.
Similarly, researchers utilize the mutual information between $\mathbf{H}$ and $\mathbf{Y}$ to characterize system performance~\cite{zhang2021overview}.

However, as introduced in Sec.~\ref{subsec:sensing_systems}, we are increasingly unsatisfied with merely sensing the channel response $\mathbf{H}$ and ubiquitously sensing with communication devices.
In this case, mutual information $I(\mathbf{Y};\mathbf{H})$ cannot fully characterize the sensing performance of the system.
For instance, MapFi~\cite{tong2021mapfi} shows that under the same estimation level for $\mathbf{H}$, the accuracy of localization using angle of arrival varies with different orientations of the antenna array.
Moreover, traditional channel theory is often based on Shannon's second law, which assumes that the random variables used for encoding are independent and identically distributed, a condition difficult to meet when conducting ubiquitous sensing.
Therefore, we need to construct a new channel model to adapt to the increasingly developed integrated communication and sensing systems.

%% file: sections/conclusion.tex
\section{Conclusion}
\label{sec:conclusion}
In this paper, we establish a channel model suitable for ubiquitous sensing, where we associate the sensing task with the received channel embedding through discrete task mutual information.
Compared to the sensing mutual information in the integrated sensing and communication system, discrete task mutual information can more accurately evaluate the performance of the sensing system.
Unlike traditional communication channel models, in sensing channels, it is difficult to maintain the independent and identically distributed characteristics among different random variables.
For discrete task sensing channels, we provide upper and lower bounds for the expected error of sensing based on discrete task mutual information, and give a sufficient condition for achieving lossless sensing.
We conduct case studies on four common sensing applications based on experimental data and simulation data.
The results show that discrete task mutual information has a strong similarity with sensing accuracy. This provides a theoretical evaluation method for the performance of integrated sensing and communication systems beyond experimental evaluation.